\documentclass[lettersize,journal]{IEEEtran}
\usepackage{amsmath,amsfonts}
\usepackage{algorithmic}
\usepackage{array}
\usepackage{tabularx}
\usepackage{booktabs}
\usepackage{cite}
\usepackage{color,soul}
\usepackage{textcomp}
\usepackage{stfloats}
\usepackage{url}
\usepackage{verbatim}
\usepackage{graphicx}
\usepackage[export]{adjustbox}
\usepackage{hyperref}
\hypersetup{colorlinks,allcolors=black}
\usepackage{balance}
\usepackage[normalem]{ulem}
\usepackage[user]{zref}
\usepackage[table]{xcolor}
\usepackage{tabularx,booktabs,array}

\definecolor{colA}{HTML}{E8F0FA} 
\definecolor{colB}{HTML}{E8F5EF} 
\definecolor{colC}{HTML}{FFF1DE} 
\definecolor{colD}{HTML}{FBEAEC} 
\definecolor{colE}{HTML}{F0EBFA} 

\newcolumntype{A}{>{\columncolor{colA}\raggedright\arraybackslash}p{2.35cm}}
\newcolumntype{B}{>{\columncolor{colB}\raggedright\arraybackslash}X}
\newcolumntype{C}{>{\columncolor{colC}\raggedright\arraybackslash}X}
\newcolumntype{D}{>{\columncolor{colD}\raggedright\arraybackslash}X}
\newcolumntype{E}{>{\columncolor{colE}\raggedright\arraybackslash}p{2.35cm}}

\newcommand{\sot}[1]{}

\newcounter{revc}
\makeatletter \zref@newprop{revcontent}{} \zref@addprop{main}{revcontent}
\zref@newprop{revsec}{} \zref@addprop{main}{revsec}
\zref@newprop{revpage}{} \zref@addprop{main}{revpage}
\newcommand{\revi}[2]{\zref@setcurrent{revsec}{\thesection}\zref@setcurrent{revpage}{\thepage}%
\zref@setcurrent{revcontent}{#2}\refstepcounter{revc}\label{#1}\zlabel{#1}\textcolor{blue}{#2}}
\makeatother

\newcommand{\figplaceholder}[3]{%
  \IfFileExists{figures/#1.pdf}{\includegraphics[width=#2]{figures/#1.pdf}}{%
  \IfFileExists{figures/#1.png}{\includegraphics[width=#2]{figures/#1.png}}{%
    \fbox{\parbox[c][#3][c]{\dimexpr#2-2\fboxsep-2\fboxrule\relax}{\centering\small\textbf{Placeholder:} \texttt{\detokenize{figures/#1.pdf}} or \texttt{.png}\\[2pt]
    Generate with the prompt in \texttt{figure\_prompts.md}.}}}}}

\begin{document}

\title{Full-Duplex-Enabled Integrated Sensing and Communications: Opportunities and Challenges}

\author{Chandan Kumar Sheemar, George C. Alexandropoulos, Sourabh Solanki,  \\
Eva Lagunas, Jorge Querol, and Symeon Chatzinotas
\thanks{C. K. Sheemar, E. Lagunas, J. Querol, and S. Chatzinotas are with the Interdisciplinary Centre for Security, Reliability and Trust (SnT), University of Luxembourg (email: \{name.surname\}@uni.lu).} 
\thanks{G. C. Alexandropoulos is with the National and Kapodistrian University of Athens, Greece and the University of Illinois Chicago, USA (email: alexandg@di.uoa.gr).} 
\thanks{S. Solanki is with the National Institute of Technology Warangal, India (email:ssolanki@nitw.ac.in).}
}

\maketitle

\begin{abstract}
Integrated sensing and communications (ISAC) is one of the six usage scenarios of IMT-2030, with in-band full duplex (IBFD) being the most direct way to realize it. In this paradigm, a base station that transmits and receives on the same time-frequency resources can sense continuously with its own downlink waveform while serving uplink traffic. This article revisits full-duplex-enabled ISAC (FD-ISAC) in light of the technologies lately shaping sixth-generation (6G) radio design, such as extremely large multiple-input multiple-output (XL-MIMO) systems operating in the near field, holographic and dynamic metasurface apertures, fluid and movable antennas, beyond-diagonal reconfigucable intelligent surfaces (BS-RISs) and wave-domain processing with stacked intelligent metasurfaces. We present a holistic FD-ISAC architecture built around a multi-stage, echo-preserving self-interference cancellation chain, and map each 6G technological enabler to the gains it brings and the risks it introduces. A case study with a $28$~GHz full-duplex XL-MIMO base station showcases that near-field-aware beamforming combined with wave-domain, analog, and digital cancellation delivers more than twice the throughput of half-duplex and 3GPP sub-band full-duplex baselines at the same localization-accuracy bound. 
It is also demonstrated that incorporating a group-connected BD-RIS recovers most of the fully-connected gain at a small fraction of its circuit complexity. The article concludes discussing emerging use cases and open challenges.
\end{abstract}

\section{Introduction}
Sixth-generation (6G) wireless networks are expected to merge communication, localization, and environmental perception into a single radio infrastructure. The ITU-R IMT-2030 framework lists integrated sensing and communications (ISAC) among its six usage scenarios~\cite{ITU2023M2160}, and a large body of work has established the relevant waveform, beamforming, and information-theoretic foundations~\cite{10979960}. Reusing the communication waveform, spectrum, and hardware for sensing promises higher spectral and energy efficiency than deploying separate radars, and gives the network native awareness of user equipment (UE), objects, and the propagation environment.

The most natural ISAC node is a base station (BS) that senses through the echoes of its own downlink (DL) signal. This monostatic mode requires receiving while transmitting on the same band, which is exactly what in-band full
duplex (IBFD) radios do~\cite{Barneto2021WCM}. FD-ISAC therefore removes the time gaps of half-duplex (HD) radar operation, senses continuously while carrying DL and uplink (UL) traffic, and reduces latency for control loops that depend on fresh sensing data. It also differs fundamentally from conventional FD communications: the self-interference (SI) can be up to $100$~dB stronger than the desired signals, yet the sensing echo is itself a delayed, attenuated copy of the transmitted waveform~\cite{Smida2024SPM}. Cancellation must therefore suppress SI without erasing the echo, within the dynamic range of practical transceivers~\cite{Sheemar2022OJCOMS}.

Industry has taken a first step: 3GPP Release~18 studied sub-band full duplex (SBFD), in which a time-division duplexing (TDD) gNB receives an UL sub-band while transmitting on adjacent DL sub-bands~\cite{3GPP38858}. SBFD validates panel isolation and cross-link interference (CLI) handling at network scale, but it keeps UL and DL apart in frequency and therefore does not, by itself, cannot provide in-band monostatic sensing. At the same time, the assumptions under which FD-ISAC was first studied are changing. Antenna arrays are becoming electrically huge, which places UEs and targets in the radiative near field~\cite{Cong2024NFISAC}. Apertures are becoming extremely large (XL)~\cite{FD-XL-MIMO} and/or holographic~\cite{Sheemar2025JSAC}, antennas can change position~\cite{New2025COMST}, 
reconfigurable intelligent surfaces (RISs) are evolving from phase-only reflectors into interconnected scattering networks~\cite{Li2024ComMag}, and stacked metasurfaces can perform signal processing directly on the wave domain~\cite{An2023JSAC}. Each of the latter brings new degrees of freedom for FD-ISAC, but also introduces new failure modes. This article provides the following four contributions:
\begin{itemize}
\item We explain why legacy FD-ISAC models break down for 6G wireless networks and summarize, in Table~\ref{tab:enablers}, what each technology enabler offers and which are its risks; the latter are evaluated through representative simulations.
\item We propose a holistic FD-ISAC architecture with a multi-stage, echo-preserving SI cancellation (SIC) chain that combines wave-domain, analog, and digital processing.
\item We present a case study against baselines, namely HD TDD, 3GPP SBFD, digital-only SIC, as well as far-field beamforming design, together with ablations on the analog-to-digital converter (ADC) resolution and consideration of a supporting beyond diagonal RIS (BD-RIS).
\item We outline use cases and open challenges, from echo-preservation metrics to standardization.
\end{itemize}

\section{Why 6G Changes the FD-ISAC Paradigm?}
Emerging 6G physical-layer technologies introduce new opportunities as well as challenges for FD-ISAC, as summarized in Table~\ref{tab:enablers} and discussed in the following.

\subsection{Echoes Hide Under Self-Interference}
In FD-ISAC systems, the SI and target echoes are mixed, so a canceller that blindly learns whichever signal correlates with the transmitted one also removes sensing information from nearby targets. The problem is compounded by hardware: transmitter distortion travels through the SI channel but is unknown to the digital canceller, and the ADCs quantize the full SI before any digital processing, raising the noise floor in proportion to the residual SI. This limited-dynamic-range effect is well documented for millimeter-wave FD transceivers with hybrid beamforming~\cite{Sheemar2022OJCOMS}. The SI is also not a single path: besides the direct coupling between transmit and receive panels, it contains reflections from nearby structures, which are environment-dependent, frequency-selective over hundreds of megahertz, and time-varying, and which overlap in delay with the echoes of close targets. Phase noise and power-amplifier nonlinearity add components that a linear canceller cannot model~\cite{9789966}. XL-MIMO makes it more pressing, because hundreds of receive radio-frequency (RF) chains push designers toward low resolution ADCs. Reduced-complexity multi-tap analog cancellation~\cite{Smida2024SPM} and RISs that reshape the SI channel~\cite{Sheemar2023RISFD} show that cancellation must be co-designed with beamforming and sensing rather than appended as a final filter.
\begin{table*}[!t]
\centering
\caption{Key 6G enablers for FD-ISAC (SIM: stacked intelligent metasurfaces, DoF: degrees of freedom, CSI: channel state information).}
\label{tab:enablers}
\scriptsize
\setlength{\tabcolsep}{3.5pt}
\renewcommand{\arraystretch}{1.12}
\begin{tabularx}{\textwidth}{@{}ABCDE@{}}
\toprule
\textbf{Technology Enabler} &
\textbf{New DoF} &
\textbf{FD-ISAC benefits} &
\textbf{Key challenges} &
\textbf{Evaluation} \\
\midrule

\textbf{Near-field XL-MIMO}
\cite{Cong2024NFISAC,FD-XL-MIMO}
&
Joint range--angle focusing and nulling
&
Co-angle UE/target separation; range-aware sensing; spatial SI/CLI suppression
&
Spatial non-stationarity; aperture calibration; near-field SI coupling
&
Figs.~\ref{fig:nf}--\ref{fig:tradeoff}: $512$ elements, $2.74$\,m
\\

\textbf{Holographic MIMO}
\cite{Sheemar2025JSAC}
&
Dense programmable apertures with few radio-frequency chains
&
Energy-efficient focusing; Cram\'er-Rao-bound-aware sensing; compact FD arrays
&
Limited digital SIC DoF; mutual coupling; hardware constraints
&
Architecture: Fig.~\ref{fig:arch}
\\

\textbf{Fluid/movable antennas}
\cite{New2025COMST}
&
Reconfigurable antenna positions
&
UE-side CLI avoidance; spatial diversity; bistatic sensing
&
CSI acquisition; switching delay; mobility-aware sensing
&
Fig.~\ref{fig:tradeoff}: $12$\,dB CLI reduction
\\

\textbf{BD-RISs}
\cite{Li2024ComMag}
&
Interconnected elements; non-diagonal scattering
&
Joint coverage extension and sensing; flexible SI-path shaping~\cite{Chepuri2023SPM,Sheemar2023RISFD}
&
Circuit complexity; coupling; control overhead
&
Fig.~\ref{fig:enablers}(b)
\\

\textbf{SIM enabling wave-domain computing}
\cite{An2023JSAC}
&
Cascaded programmable wave transforms
&
Pre-ADC SI suppression; analog feature extraction
&
Insertion loss; calibration; bandwidth constraints
&
Fig.~\ref{fig:enablers}(a)
\\

\textbf{Multi-stage SIC}
\cite{Barneto2021WCM,Smida2024SPM,Sheemar2022OJCOMS}
&
Passive, analog, and digital cancellation
&
transmit-distortion mitigation; ADC protection; continuous sensing
&
Echo preservation; wideband and time-varying SI
&
Figs.~\ref{fig:tradeoff}--\ref{fig:enablers}(a)
\\

\textbf{3GPP's SBFD}
\cite{3GPP38858}
&
UL/DL frequency partition within TDD
&
Improved UL coverage, latency, and practical deployment
&
UE--UE CLI; no in-band monostatic sensing
&
Fig.~\ref{fig:tradeoff}
\\

\bottomrule
\end{tabularx}
\vspace{-2mm}
\end{table*}
\subsection{Near Field is the Default Regime with XL-MIMO}
The radiative near field extends to roughly twice the squared aperture divided by the wavelength. A $2.7$~m array at $28$~GHz is in the near field up to about $1.4$~km, so practically every UE and target of a small cell lies inside it. Spherical wavefronts let the BS focus energy on a location rather than a direction, separating UEs and targets that share an angle but differ in range; this enables ``location-division'' multiplexing and sensing~\cite{Cong2024NFISAC}. Focusing in range is, however, only sharp well inside the near field: the depth of focus grows roughly with the square of the distance, so range selectivity is strongest for nearby UEs and targets, and fades toward the Rayleigh distance. Near-field operation also creates spatial non-stationarity across the aperture, where different UEs see different parts of the array, and makes the SI coupling between the transmit and receive panels itself a near-field channel that can be shaped with spherical-wave nulling. Initial studies on FD holographic MIMO for near-field ISAC confirm that SIC and analog/digital beamformers must be optimized jointly~\cite{FD-XL-MIMO}.

\subsection{Apertures Become Reconfigurable}
Holographic MIMO, realized with reconfigurable holographic surfaces or metasurface antennas, packs many low-power radiating elements next to few RF chains and allows fine wavefront control; Cram\'er-Rao-bound (CRB) optimized holographic beamforming can trade communication rate against three-dimensional sensing accuracy~\cite{Sheemar2025JSAC}. Because such apertures expose only a few digital ports, most of the SI suppression must happen before signal converters, which raises the importance of analog- and wave-domain cancellation. At the UE side, fluid antenna systems (FAS) and movable antennas~\cite{New2025COMST} add a new control knob: the antenna \emph{position}. By moving the radiating point within a few wavelengths, a device can exploit spatial diversity, or deliberately sit in a deep fade of an interfering channel, with a single RF chain. Pinching antennas on dielectric waveguides extend the same idea to large-scale repositioning. For FD-ISAC, position control can suppress UL-to-DL CLI, the main side effect of serving UL and DL UEs simultaneously, providing additional bistatic viewpoints. At the BS, movable elements could likewise be positioned to reduce panel-to-panel coupling, although switching speed and mechanical reliability remain open questions.

\subsection{Metasurfaces Become Network Nodes and Processors}
RIS-assisted ISAC creates virtual line-of-sight links toward blocked UEs and non-line-of-sight (NLoS) targets~\cite{Chepuri2023SPM}. The late BD-RIS architecture interconnects elements through tunable impedances, generalizing the diagonal phase-shift model to single-, group-, and fully-connected architectures, and matching the circuit physics more closely~\cite{Li2024ComMag}. This additional reconfigurability lets a single metasurface split power among several beams without the losses of phase-only control, and supports multi-sector modes that extend coverage to the full space around the surface. The price is circuit complexity, which grows quadratically with the group size. Stacked intelligent metasurfaces (SIM) go one step further: cascaded programmable layers implement transformations on the wave itself, performing transmit precoding and receive combining in the wave domain with few RF chains~\cite{An2023JSAC}. For FD-ISAC, this opens a new possibility: projecting the SI out \emph{before} it reaches the ADCs, while letting the echoes pass through.

\subsection{FD in Standards and Relevance to ISAC}
SBFD is the first standardized form of a network-side FD system~\cite{3GPP38858}. Its evaluation methodology, with gNB-side operation, legacy and SBFD-aware UE coexisting, and explicit CLI measurement and coordination is a valuable template for FD-ISAC. However, SBFD separates the UL sub-band from DL transmissions, so monostatic sensing still needs either dedicated sensing time or an IBFD-capable receiver. The case study in Section~\ref{sec:case} quantifies this gap.

\section{A Holistic FD-ISAC Architecture for 6G}

\begin{figure*}[!t]
\centering
\includegraphics[width=\linewidth]{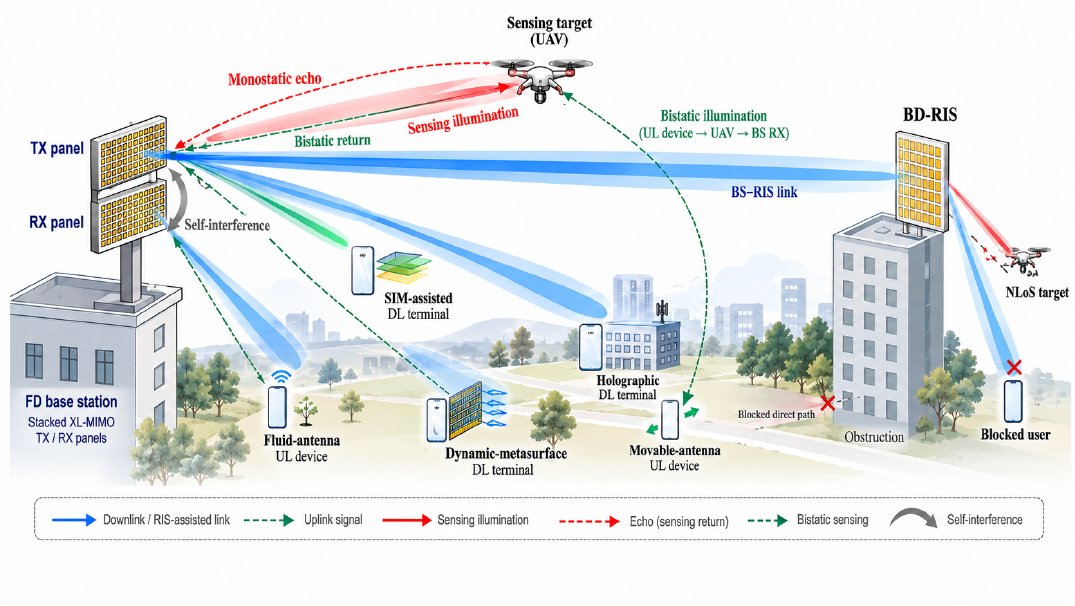}
\caption{The proposed holistic 6G FD-ISAC architecture. An FD base station with stacked XL-MIMO transmit and receive panels (possibly holographic) serves DL and UL users while sensing targets monostatically. UL devices with fluid antennas avoid CLI toward DL users and provide bistatic views; a BD-RIS serves a blocked user and illuminates an NLoS target; a wave-domain processor and analog canceller protect the dynamic range of the FD-ISAC's receiver.}
\label{fig:arch}
\end{figure*}

Figure~\ref{fig:arch} depicts the proposed FD-ISAC architecture. An FD BS with stacked XL-MIMO transmit and receive panels serves multiple DL and UL UEs on the same time-frequency resources, while localizing and tracking targets with the echoes of its own transmissions~\cite{FD-XL-MIMO}. The signal at the BS receiver is a superposition of: \textit{i}) residual SI after all cancellation stages, \textit{ii}) multi-UE UL signals, \textit{iii}) monostatic echoes of the DL and sensing waveforms, \textit{iv}) reflections of UL signals from targets and clutter, which act as additional bistatic sensing views, and \textit{v}) external interference and noise. The architecture treats components \textit{ii})--\textit{iv}) as information to be preserved and only component \textit{i}) as a nuisance to be removed.

\subsection{Multi-Stage, Echo-Preserving SIC}
\begin{figure*}[!t]
\centering
\includegraphics[width=\linewidth]{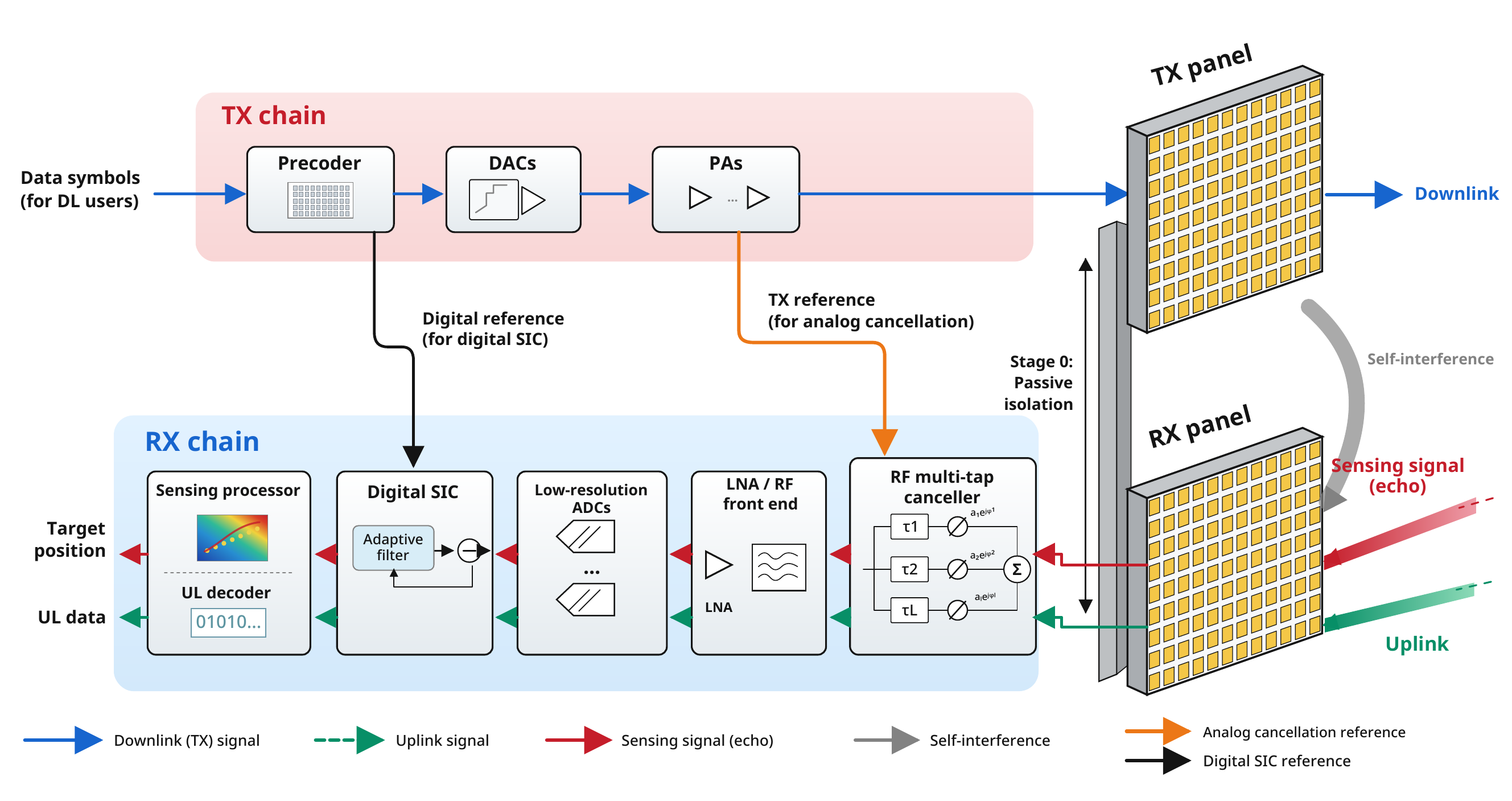}
\caption{Multi-stage, echo-preserving SIC chain of the FD-ISAC receiver: passive isolation, wave-domain spatial projection of the SI subspace, RF multi-tap cancellation of transmitter distortion, low resolution ADCs, digital SIC of the known linear SI, and a sensing processor that separates targets from clutter (LNA: low noise amplifier).}
\label{fig:sic}
\end{figure*}

Figure~\ref{fig:sic} illustrates the receive chain of the FD-ISAC architecture. Stage~0 is passive isolation from panel separation, shielding, and element patterns. Stage~1 is a wave-domain spatial projector, realizable with a SIM, a BD-RIS-type network, or a hybrid analog combiner. Since SI generated by a beamformed transmitter occupies a subspace whose dimension equals the number of transmitted streams, a few receive dimensions suffice to remove it before the ADCs; target echoes and UL UEs, which arrive through different near-field channels, pass almost untouched. Stage~2 is a multi-tap RF canceller fed from the power amplifier outputs. It removes what spatial projection cannot: transmitter distortion, which is spread over all receive dimensions. Stage~3 is the bank of (possibly low resolution) ADCs, and Stage~4 is digital cancellation of the known linear SI. Finally, the sensing processor separates static clutter from targets with structure-aware estimators, so that the SI canceller is never allowed to absorb the echoes it is meant to reveal. The stages are complementary, and the following case study shows that dropping either the wave-domain or the RF stage costs either ADC(s) resolution or sensing accuracy.

\subsection{Location-Aware Transmission}
With near-field operation, the DL precoder and the sensing beam are designed on locations and regions rather than directions. The sensing beam is focused on the predicted target position, available from tracking, or a region of interest, and placed in the null space of the DL UEs' channels, so that a UE sharing the target's angle but not its range sees almost no sensing power. Receive processing should be designed jointly so that the spatial SI projection preserves the echo subspace, including the directions that carry position information. In our case study, a plain projection onto the complement of the SI subspace already keeps virtually all of the echo energy.

\subsection{Reconfigurable Devices and RISs}
UL devices with FAS select, among a few tens of ports spread over a couple of wavelengths, the port that minimizes their aggregate CLI towards co-scheduled DL UEs. With line-of-sight links toward the BS, this costs essentially nothing in UL gain. A BD-RIS mounted on a building facade extends both services around blockages: it splits the incident beam between a blocked user and an NLoS target with far less loss than a phase-only-adjustable RIS, and group-connected architectures keep the number of tunable circuit elements manageable. Metasurfaces and UL devices also provide bistatic sensing views that BS fuses with its monostatic measurements.

\subsection{Bistatic and Multi-Static Views for Free}
In FD-ISAC, the UL is not only traffic. Once UL symbols are decoded, their reflections from targets become known-waveform bistatic measurements at the BS, captured at the same time as the monostatic echoes and without extra transmissions. Because the BS already estimates UL timing and frequency offsets, the synchronization burden of bistatic sensing is largely absorbed by the UL procedures. Metasurfaces add further views: an RIS illuminating an NLoS target returns its echo through the same reciprocal surface, and UL devices that reposition their fluid antennas over time effectively provide snapshots from slightly different viewpoints. Fusing these views is where multi-static gains come from, and where the hardest synchronization and calibration problems lie.

\subsection{Control and Signaling}
FD-ISAC scheduling must jointly allocate DL/UL UEs, sensing beams, FAS ports, and metasurface configurations under SI and CLI constraints. A practical split is a slow loop for RIS and isolation configuration, and a fast loop for beams, power, and ports driven by tracking predictions and CLI measurements of the kind SBFD already standardizes.

\section{Case Study: Near-Field FD-ISAC With Reconfigurable Front Ends}
\label{sec:case}

\subsection{Setup and Baselines}
We consider a $28$~GHz FD BS with $400$~MHz bandwidth and two stacked $512$-element half-wavelength uniform linear arrays ($2.74$~m aperture, $10$~cm vertical separation, $55$~dB passive isolation, and $24$~dBm transmit power). The Rayleigh distance is about $1.4$~km. The BS serves four DL UEs and two UL UEs ($20$~dBm) at $15$--$55$~m, and localizes a small unmanned aerial vehicle (UAV) (radar cross-section of $-20$~dBsm) at $35$~m that shares its angle with DL UE~$1$ at $15$~m. BS links use exact spherical-wave line-of-sight channels; UE--UE CLI follows Rayleigh fading with 3GPP urban-microcell NLoS path loss. UL UEs carry a $32$-port FAS over two wavelengths. Hardware follows the limited-dynamic-range model~\cite{Sheemar2022OJCOMS} with $-40$~dB transmitter distortion and ADC noise set by its resolution ($8$ bits by default) with $10$~dB back-off. The wave-domain and RF stages provide $30$~dB and $25$~dB of suppression, respectively. Communication performance is the sum UL and DL throughput with regularized zero-forcing precoding and minimum-mean-square-error successive-interference-cancellation detection. Sensing accuracy is the root-CRB (RCRB) on the two-dimensional target position, computed over $32$ sub-bands and one $14$-symbol slot; this captures both delay and wavefront-curvature information. Note that sweeping power split between communication and sensing beams traces the trade-off.

We compare against: \textit{i}) an ideal IBFD bound with perfect SIC and no CLI; \textit{ii}) the proposed scheme with fixed-position UE antennas; \textit{iii}) IBFD with far-field (angle-only) beam design; \textit{iv}) IBFD with digital-only SIC; \textit{v}) 3GPP-style SBFD with a $20\%$ UL sub-band, $10\%$ guard bands, the $1$~dB SI desensitization assumed in 3GPP evaluations~\cite{3GPP38858}, and $20$~dB adjacent-channel CLI attenuation; and \textit{vi}) HD-TDD with a $3:1$ DL:UL split. Baselines \textit{v}) and \textit{vi}) sense in dedicated slots that we idealize as free of SI, which favors them.

\subsection{Near-Field Focusing Resolves Co-Angle Users and Targets}
\begin{figure*}[!t]
\centering
\includegraphics[width=\textwidth]{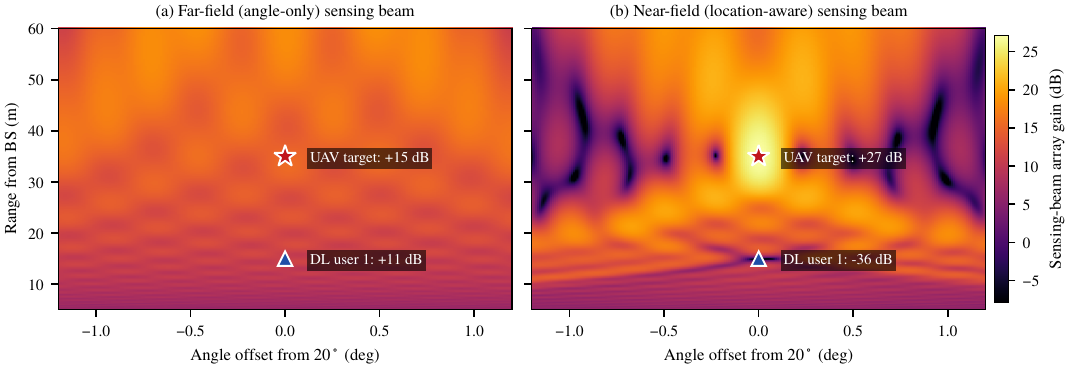}
\caption{Array gain of the sensing beam over a range-angle window centered on the direction shared by DL user~$1$ ($15$~m) and a UAV target ($35$~m): (a) far-field (angle-only) design; (b) near-field design focused on the UAV with a null on the user. A $512$-element antenna array at the transmitter and receiver of the FD-ISAC system at $28$~GHz has been considered with $60\%$ of its transmit power allocated on communication beams.}
\label{fig:nf}
\end{figure*}

Figure~\ref{fig:nf} shows the sensing beam around the shared direction. Under the far-field model, DL UE~$1$ and the UAV are indistinguishable (channel correlation equal to one), so the sensing beam either leaks into the UE or is lost. Here, it delivers $+11$~dB of array gain at the UE and only $+15$~dB at the UAV, because a planar-wave beam is defocused inside the near field. Near-field design reduces the correlation to $0.19$, places a $-36$~dB null on the UE, and concentrates $+27$~dB on the UAV: a $47$~dB swing in leakage and $12$~dB more illumination from the same aperture.

\subsection{Throughput--Accuracy Trade-Off}
\begin{figure}[!t]
\centering
\includegraphics[width=\columnwidth]{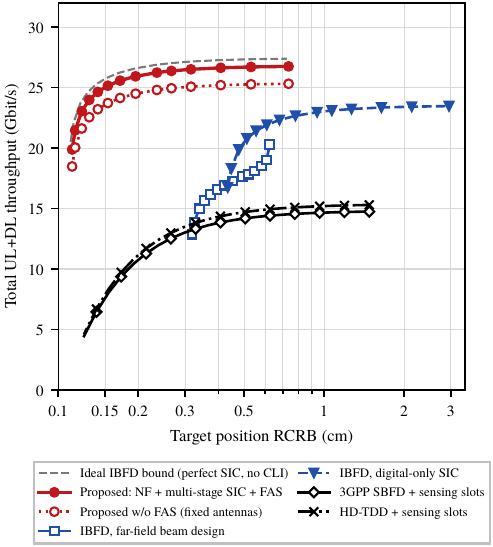}
\caption{Sum UL and DL throughput vs. target-position RCRB. IBFD curves sweep the power split between communication and sensing beams; SBFD and HD-TDD sweep the fraction of time spent in (idealized, SI-free) sensing slots.} 
\label{fig:tradeoff}
\end{figure}

As reported in Fig.~\ref{fig:tradeoff}, the proposed design stays within about $0.6$~Gbit/s ($2.5\%$) of the ideal IBFD bound over the entire trade-off. At an RCRB of $2$~mm, it delivers $26$~Gbit/s, compared to $10.7$ and $11.1$~Gbit/s for SBFD and HD-TDD, respectively, that is, $2.3$ to $2.4$ times more,\footnote{The factor-of-two limit of full duplex applies to communication throughput alone. Without a sensing requirement, the gain here is about $1.75$ ($26.8$ vs.\ $15.3$~Gbit/s), i.e., below two, because the DL and UL rates differ and HD-TDD already gives three quarters of its time to the stronger DL. At a fixed RCRB, however, the comparison also accounts for the cost of sensing: to reach $2$~mm, HD-TDD and SBFD must reserve about $27\%$ of each slot for dedicated sensing, because they cannot collect echoes while transmitting, whereas IBFD senses with its own DL waveform and gives up only about $3\%$ of its throughput to the sensing beam. Combining the duplexing gain ($1.75$), the sensing time saved ($1/0.73 \approx 1.37$), and the small power-split loss ($0.97$) gives about $2.3$, as in Fig.~\ref{fig:tradeoff}; SBFD follows the same reasoning.} and its peak throughput is $26.8$~Gbit/s compared to $15.3$~Gbit/s for HD-TDD. The performance with SBFD and HD-TDD is almost indistinguishable: SBFD re-partitions the carrier in frequency instead of time, which helps UL coverage and latency, but neither adds resources nor enables in-band sensing. The far-field design cannot push the RCRB below $3.2$~mm and loses $34\%$ of the throughput at $5$~mm, because the sensing beam jams the co-angle UE. Digital-only SIC floors at $4.3$~mm, almost four times worse than the proposed design's best accuracy, since transmitter distortion and quantization noise raise the receiver floor. Finally, FAS lowers the CLI at DL UE~$1$ by $12$~dB, from $10$~dB above the noise floor to slightly below it, worth $1.4$~Gbit/s with a single RF chain per device.

\subsection{The Role of 6G Technological Enablers}
\begin{figure*}[!t]
\centering
\includegraphics[width=\textwidth]{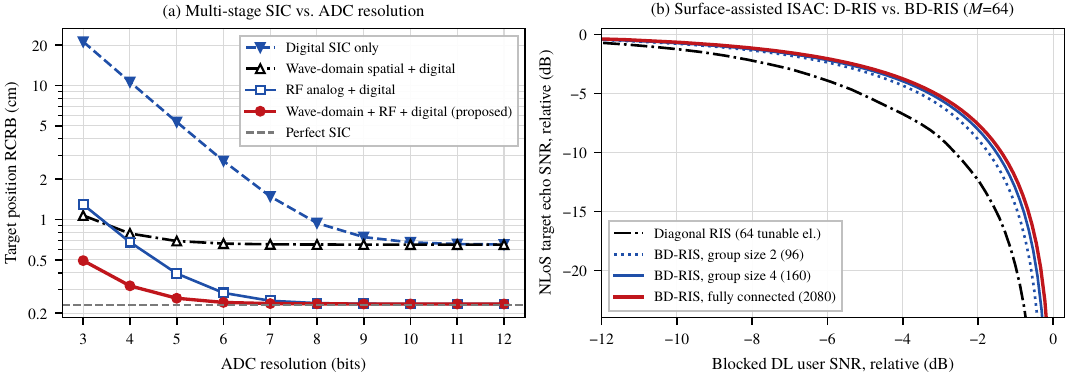}
\caption{Contribution of individual enablers. (a) Target-position RCRB vs. ADC resolution for different SIC chains ($30$~dB wave-domain and $25$ allocated on communication beams); (b) Trade-off between blocked-user SNR and NLoS-target echo SNR, relative to the single-function optimum, for diagonal RIS and group-/fully-connected BD-RIS (Rician user link, $100$ channel realizations; the number of BD-RIS tunable circuit elements is given in parentheses).}
\label{fig:enablers}
\end{figure*}

Figure~\ref{fig:enablers}(a) isolates the SIC chain of the proposed FD-ISAC structure. As observed, with digital SIC alone, the RCRB grows from $6.5$~mm at $12$-bit ADCs to $10.6$~cm at $4$ bits, and never falls below $2.8$ times the perfect-SIC value, because transmitter distortion cannot be cancelled digitally. Wave-domain projection alone hits the same floor for the same reason. The RF stage removes it but still needs about $8$ bits. Combining both reaches within $5\%$ of perfect SIC with $6$-bit ADCs and keeps $3.2$~mm accuracy and $23.9$~bit/s/Hz of UL spectral efficiency at $4$ bits, where digital-only SIC collapses to $4.5$~bit/s/Hz. The projection removes less than $0.001$~dB of echo energy: it is echo-preserving by construction.

Figure~\ref{fig:enablers}(b) shows a $64$-element RIS that serves a blocked UE and illuminates an NLoS UAV. At a $2$~dB loss in UE signal-to-noise ratio (SNR), a diagonal RIS retains an echo SNR $12.3$~dB below the sensing-only optimum, whereas a fully-connected BD-RIS retains $-7.6$~dB, i.e., $4.7$~dB more. The gain is doubled by the round trip through the metasurface. A group-connected BD-RIS with groups of four comes within $0.5$~dB of the fully-connected design using $160$ instead of $2080$ tunable circuit elements, a practical sweet spot.

\subsection{Lessons Learned and Limitations}
Four lessons emerge from the previous investigation. First, in-band operation is what matters for ISAC: SBFD helps the UL, but only IBFD lets every DL symbol double as a sensing probe. Second, near-field awareness is not optional with XL arrays, because angle-only design either blinds the radar or jams co-angle UEs. Third, the SIC chain must be split across domains according to the rank of what it removes: spatial projection for the low-rank SI signal, analog/RF cancellation for full-rank distortion, and digital processing for the known remainder. Fourth, reconfigurability pays off where it targets a dominant impairment, such as CLI for FAS and multi-beam loss for BD-RIS, at moderate hardware cost. The study assumed perfect channel knowledge, a single point target without clutter, ideal sensing slots for the HD and SBFD baselines, and a fully-digital baseband behind the wave-domain stage; extending it to clutter-rich, imperfect-CSI, and hybrid beamforming conditions is part of the open agenda below.

\section{Emerging FD-ISAC Use Cases}
\textbf{Low-Altitude Airspace:} Small UAVs have tiny radar cross-sections and fly close to dense UE populations. Continuous FD monostatic sensing with near-field focusing, as in our case study, can track them without pausing communications, while RISs extend coverage behind buildings. The same infrastructure can enforce flight corridors and serve the UAVs' own command-and-control links.

\textbf{Industrial Digital Twins:} Factories need persistent tracking of robots, vehicles, and people next to latency-critical control traffic. FD-ISAC delivers both on one carrier, and location-aware beams keep sensing energy away from sensitive equipment. Short ranges and metallic clutter are the elements where near-field focusing and echo-preserving SIC matter most.

\textbf{Cooperative Perception for Mobility:} Roadside units can exchange vehicle-to-everything messages while sensing vulnerable road UEs; UL transmissions of vehicles provide additional bistatic views that infrastructure can fuse with its own echoes, reducing the blind spots of on-board sensors.

\textbf{Extended Reality and Human-Centric Sensing:} This paradigm needs high-rate links and fine pose tracking under tight motion-to-photon latency. FD-ISAC can combine these functionalities on the same resources, while FAS devices and metasurfaces can overcome body blockage.

\textbf{Secure and Private Spaces:} Location-aware beams and RISs can create ``sensing fences'' and quiet zones in which neither data nor sensing energy is delivered, and sensing can locate suspicious emitters to support physical-layer security.

\section{Open Challenges}

\subsection{Echo-Preserving SI Cancellation and Metrics}
Cancellers are still judged by suppression depth alone. FD-ISAC needs metrics that also count what the canceller must \emph{not} remove, e.g., echo-energy or Fisher-information preservation, and estimators that separate SI, static clutter, and slow targets that all correlate with the transmitted waveform. Learning-based cancellers must be constrained accordingly, and their training data must include targets, not only SI. Co-designing waveforms, for example, by reserving resources whose echoes are easier to separate from SI, is a complementary direction.

\subsection{Practical Wave-Domain Signal Processing}
Our results assume $30$~dB of wave-domain suppression over $400$~MHz. Achieving it requires low-loss, calibrated SIM, or BD-RIS hardware that is fast enough to follow beam updates, with a stable response across the band and over temperature. Joint design with the analog canceller, which covers the full-rank distortion that spatial projection cannot, is largely open, as is the question of how many RF taps and metasurface layers are needed for a given bandwidth and array size.

\subsection{Near-Field Model Fidelity and Calibration}
Meter-scale apertures face spatial non-stationarity, partial visibility, wideband beam squint, and mutual coupling. The panel-to-panel SI channel is itself a near-field channel that drifts with temperature and mechanical stress. Low-overhead calibration and tracking of all these channels are prerequisites for the gains shown in this article, and channel models validated by near-field FD measurements are still scarce.

\subsection{Reconfigurable Antennas at Network Scale}
FAS port selection against CLI requires UE--UE channel knowledge, which today exists only as coarse CLI measurements. Signaling, switching latency, and the interplay between port changes and sensing with time-varying geometry need protocol-level answers, as does the question of whether port selection should be UE-driven or network-controlled.

\subsection{RIS Control and Modeling}
BD-RIS optimization necessitates physics-consistent scattering models with coupling and losses, wideband designs, and CSI estimation beyond diagonal phase shifts. Near an FD BS, an metasurface can also create new SI paths. Jointly shaping service, sensing, and SI is an open design problem, and so is finding BD-RIS architectures between group- and fully-connected that keep most of the gain at low cost.

\subsection{Multi-Static Fusion, Security, and Privacy}
Combining monostatic, bistatic, and RIS-assisted views requires time, frequency, and phase synchronization across nodes, and feature-level fusion under practical fronthaul limits. Sensing-capable networks also expand the attack metasurface: sensing leakage, spoofed targets, and tampered surface configurations. Regulation may demand sensing transparency and enforceable privacy zones, which in turn requires beam and metasurface control that can be audited.

\subsection{From SBFD to In-Band FD-ISAC in Standards}
SBFD provides the isolation requirements, CLI measurements, and coexistence framework on which FD-ISAC can build upon. The next steps are test procedures for in-band sensing receivers, sensing-aware scheduling and reporting, and coexistence rules for UEs with reconfigurable antennas.

\section{Conclusion}
Full-duplex ISAC is moving from a conceptual 6G pillar to an engineering problem shaped by near-field propagation, dynamic-range limits, and modern reconfigurable hardware. In this article, we argued that the defining requirement is echo-preserving SI cancellation, and that 6G enablers change both how it can be achieved and what FD-ISAC can deliver. XL-MIMO turns the near field into a resource for location-aware sensing; holographic apertures and fluid antennas add spatial control at low cost; BD-RISs and SIM evolve from reflectors into wave-domain co-processors. Our case study showcased that, used together, these enablers allow an FD BS to deliver more than twice the throughput of HD and SBFD operation at the same localization accuracy, with low resolution converters and metasurfaces of practical complexity. Turning these gains into deployments requires progress on echo-aware metrics, practical wave-domain hardware, near-field calibration, protocols for UEs with reconfigurable antennas, and a standards path from SBFD to in-band FD-ISAC.

\balance
\bibliographystyle{IEEEtran}
\bibliography{refs}

\end{document}